# BCG vaccination in infancy does not protect against COVID-19.

# Evidence from a natural experiment in Sweden.

Clément de Chaisemartin[1]*, PhD ; Luc de Chaisemartin[2,3], PhD

[1]Department of Economics, University of California, Santa Barbara, Santa Barbara, CA 93106, USA

[2]Immunology Department, APHP.Nord-Université de Paris, Bichat Hospital, Paris, France

[3]Université Paris-Saclay, Inserm, "Inflammation, microbiome, immunosurveillance", 92290, Châtenay-Malabry, France.

∗To whom correspondence should be addressed; E-mail: clementdechaisemartin@ucsb.edu, telephone: +1-805-452-9642.

**Abstract**

**Background**

The Bacille Calmette-Guérin (BCG) tuberculosis vaccine has immunity benefits against respiratory infections. Accordingly, it has been hypothesized that it may have a protective effect against COVID-19. Recent research found that countries with universal Bacillus Calmette-Guérin (BCG) childhood vaccination policies tend to be less affected by the COVID-19 pandemic. However, such ecological studies are biased by numerous confounders. Instead, this paper takes advantage of a rare nationwide natural experiment that took place in Sweden in 1975, where discontinuation of newborns BCG vaccination led to a dramatic fall of the BCG coverage rate from 92% to 2% , thus allowing us to estimate the BCG's effect without all the biases associated with cross-country comparisons.

**Methods**

Numbers of COVID-19 cases and hospitalizations were recorded for birth cohorts born just before and just after that change, representing 1,026,304 and 1,018,544 individuals, respectively. We used regression discontinuity to assess the effect of BCG vaccination on Covid-19 related outcomes. This method used on such a large population allows for a high precision that would be hard to achieve using a classical randomized controlled trial.

**Results**

The odds ratio for Covid-19 cases and Covid-19 related hospitalizations were 0·9997 (CI95: [0·8002-1·1992]) and 1·1931 (CI95: [0·7558-1·6304]), respectively. We can thus reject with 95% confidence that universal BCG vaccination reduces the number of cases by more than 20% and the number of hospitalizations by more than 24%

**Conclusions**

While the effect of a recent vaccination must be evaluated, we provide strong evidence that receiving the BCG vaccine at birth does not have a protective effect against COVID-19.

**Introduction**

The Bacille Calmette-Guérin (BCG) tuberculosis vaccine has immunity benefits against non-targeted pathogens[1], and in particular against respiratory infections caused by RNA viruses like influenza[2]. Since SARS-Cov-2 is also a single-stranded RNA virus, it has been hypothesized that differences in BCG vaccination coverage could explain the wide differences in disease burden observed between countries. A pioneering preprint paper by Miller *et al*. found that countries with universal Bacillus Calmette-Guérin (BCG) childhood vaccination policies tend to be less affected by the COVID-19 pandemic, in terms of their number of cases and deaths[3]. While unpublished, this study had a great impact and gave rise to many comments and follow-up studies (reviewed in[4]). Some published studies were able to replicate this result[5,6], but several authors underlined the important statistical flaws inherent to such ecological studies and concluded that randomized controlled trials (RCT) were necessary to address the question[4,7]. As of June 5th 2020, no less than 12 randomized controlled trials (RCTs) studying the protective effect of the BCG against COVID-19 are already registered on https://clinicaltrials.gov/. However, none has a primary completion date earlier than October 1st 2020, so these RCTs' first results will not be known until at least five or six months. With the epidemic still on the rise worldwide, and in the absence of a SARS-Cov-2 vaccine, there is an urgent need to know if BCG non-specific effects could be harnessed as a substitute prophylactic treatment.

Regression discontinuity (RD) is a method designed by social scientists to assess the effect of an exposure on an outcome. It is deemed as reliable as RCTs to tease out causality from correlation[8], and typically yields results similar to those obtained in RCTs[9,10]. In this paper, we applied this method to a rare natural experiment that took place in Sweden. Sweden currently has the 5th highest ratio of COVID-19 deaths per capita in the world. In April 1975, it stopped its newborns BCG vaccination program, leading to a dramatic drop of the BCG vaccination rate from 92% to 2% for cohorts born just before and just after the change[11]. We compared the number of COVID-19 cases, hospitalizations, and deaths per capita, for cohorts born *just before* and *just after* April 1975, representing 1,026,304 and 1,018,544 individuals, respectively. Using RD, we were able to show that those cohorts do not have different numbers of COVID-19 cases, hospitalizations or deaths per capita, with a high precision that would hardly be possible to reach with a RCT design.

**Methods**

**Data sources**

Two sources of data were used. First, COVID-19 reported cases, hospitalizations, and deaths compiled by Folkhälsomyndigheten, the Public Health Agency of Sweden, as of May 17th 2020. At that date, Sweden had 30,312 reported COVID-19 cases and 3,954 reported COVID-19-related deaths (i.e with a confirmed Covid-19 diagnosis during the past 30 days)[12]. Second, Swedish demographic data publicly available from Statistics Sweden's website were used. For the number of Covid-19 cases and hospitalizations, quaterly birth cohort (QBC) from Q1-1930 to Q4-2001 and from Q1-1930 to Q4-1991 were used, respectively. For number of reported Covid-19 death, data were grouped by three yearly birth cohorts (YBC) from YBC 1930 to YBC 1980 to ensure that all groups had at least five deaths recorded. The detail of the raw and constructed variables used in the analysis are described in Supplementary Table 1.

**Statistics**

Regression discontinuity (RD) is a commonly-used method to measure the effect of a treatment on an outcome[13]. It is applicable when only individuals that satisfy a strict criterion are eligible for a policy. Then, RD amounts to comparing the outcome of interest among individuals just above and just below the eligibility threshold. In this study, RD will amount to comparing the number of COVID-19 cases, hospitalizations, and deaths among individuals born just before and just after April 1st 1975. The effect of universal BCG vaccination for individuals born around April 1st, 1975 was estimated using the state-of-the-art estimator for RD[14]. The estimator amounts to comparing treated and control units, in a narrow window around April 1st 1975. It uses linear regressions of the outcome on birth cohort to the left and to the right of the threshold, to predict the outcome of treated and untreated units at the threshold. Then, the estimator is the difference between these two predicted values. The estimator and 95% confidence interval were computed using the rdrobust Stata command, see[15].

**Results**

This study uses the number of COVID-19 cases per 1000 inhabitants for quarterly birth cohorts born between Q1-1930 and Q4-2001, the number of COVID-19 hospitalizations per 1000 inhabitants for cohorts born between Q1-1930 and Q4-1991, and the number of COVID-19 deaths per 1000 inhabitants for groups of three yearly birth cohorts, from 1930-1931-1932 to 1978-1979-1980. These variables were constructed using data compiled by the Public Health Agency of Sweden; see the supplementary Table 1 for details.

In an RD design, the presence or absence of a treatment effect can be assessed visually, by drawing a scatter-plot with the variable determining eligibility on the x-axis, and the outcome variable on the y-axis. If one observes that the relationship between these two variables jumps discontinuously at the eligibility threshold, this is indicative of a treatment effect. Accordingly, Figure 1 shows no discontinuity in the numbers of COVID-19 cases per 1000 inhabitants for cohorts born just before and just after April 1975. This suggests that universal BCG vaccination has no effect on the number of COVID-19 cases per 1000 inhabitants for individuals born in 1975. Figures 2 and 3 show that similar conclusions apply when one looks at the number of COVID-19 hospitalizations per 1000 inhabitants and at the number of COVID-19 deaths per 1000 inhabitants. The number of deaths per 1000 inhabitants is several orders of magnitudes higher for older than for younger cohorts, so Figure 3 only presents those numbers for cohorts born after 1960 to keep the graph legible.

This visual analysis is confirmed by the statistical calculations. Table 1 reports RD estimates of the effect of the BCG vaccination policy on COVID-19 outcomes, using the RD estimator (see Methods). They show that the effects of the BCG vaccination policy on cases per 1000 inhabitants and hospitalizations per 1000 inhabitants are not statistically different from 0. Based on the confidence intervals, we can reject with 95% confidence that universal BCG vaccination reduces the number of cases per 1000 inhabitants by more than 0·409, an effect equivalent to 13% of the number of cases per 1000 inhabitants in the 1975 cohort. For the number of hospitalizations per 1000 inhabitants, the effect we can reject is equal to 15% of the number of hospitalizations per 1000 inhabitants in the 1975 cohort. For deaths per 1000 inhabitants, there is only two data points to the right of the Q2-1975 threshold. Therefore, the RD estimator cannot be computed for that outcome. Instead, we just compared the number of COVID-19 deaths per 1000 inhabitants in the 1972-1973-1974 and 1975-1976-1977 YBCs using a standard t-test, even though this method does not account for the fact those two groups differ in age, contrary to the RD method. Doing so, we find that the difference between the deaths per 1000 inhabitants of the two groups is not different from 0.

The effects in Table 1 are *intention-to-treat* effects [16]: not all Swedish residents born just before April 1975 received the BCG vaccine at birth, and some of those born just after April 1975 received it. In particular, foreign-born residents account for 27·2% of the Swedish population born in 1975 as per Statistics Sweden's data, and they were not affected by the 1975 policy. Among natives, the policy led to a drop of the vaccination rate from 92 to 2%[11]. Assuming that the BCG vaccination rate of foreign-born residents is the same just before and just after April 1975, a reasonable assumption as no other European country changed its BCG vaccination policy in 1975, the policy led to a drop in the BCG vaccination rate of 0·655 ((1-0·272)×(0·92-0·02)=0·655). Then, to convert the intention-to-treat effects in Table 1 into the effect of being vaccinated at birth, one needs to divide the intention-to-treat effects and their confidence intervals by 0·655, the change in the BCG vaccination rate at birth induced by the reform[17]. In odds ratios, the effects thus obtained are 0·9997 (IC95: [0·8002, 1·1992]) for Covid-19 cases, and 1·1931 (IC95: [0·7558,1·6304]) for Covid-19 related hospitalizations.

**Discussion**

In this study, we took advantage of a change in vaccination policy in Sweden to investigate the link between BCG vaccination in infancy and Covid-19 cases, hospitalizations and deaths, using a regression discontinuity approach.

Contrarily to most studies on the question, we compared Covid-19 cases between two very similar groups of people from the same country. This allows us to get rid of all confounders linked to cross-countries comparisons, and of confounders like sex or socio-economics status that are often present in observational studies that do not rely on a quasi-experimental design, unlike ours. Another strength of this study is its statistical precision. Since we could gather nationwide data in a country where Covid-19 rates are high, we are able to reject fairly small effects of the BCG vaccine. Achieving a comparable statistical precision in an RCT would require an unrealistically large sample. Even with a COVID-19 hospitalization rate of 0·5%, as among the elderly Swedish population, a randomized controlled trial that could reject BCG effects larger than 24% of the baseline hospitalization rate, as in our study, would require including around 15,000 participants.

While previous studies mostly addressed differences in BCG vaccination policies but did not account for differences in actual BCG coverage, we work with two populations with well documented and very different vaccination rates. The termination of the universal BCG vaccination program in Sweden had dramatic effects on the BCG coverage rate. Based on nationwide reports on the vaccination status of children below 7 sent to the National Bacteriological Laboratory in 1981, 92% of children born in 1974 got vaccinated, against 26% of those born in 1975, and less than 2% of those born between 1976 and 1980[11]. Among children born in 1975, most of the vaccinated children were born in the first quarter of the year, when the universal vaccination policy was still in place[11]. Prior to that, Sweden had already eliminated its revaccination program at 7 years of age in 1965. Sweden also stopped its revaccination program at 15 years of age in 1986, four years before the 1975 cohort turned 15 years old. Finally, Sweden stopped its vaccination program for conscripts in 1979, long before the 1975 cohort would do its military service[18]. Overall, children born before April 1st 1975 benefited from a BCG vaccination policy at birth, while children born after that did not benefit from any BCG vaccination policy.

This being said, there is a number of limitations to this work that one has to bear in mind. As in many other countries, Folkhälsomyndigheten's cases count probably underestimates the true number of cases, because it only includes cases confirmed by a laboratory test. Its deaths count probably underestimates the true number of COVID-19 deaths as well, as it only includes deaths where a COVID-19 diagnosis has been confirmed during the past 30 days. However, these limitations are common to all studies on this question and are unlikely to affect much the results.

Furthermore, RD estimates only apply to units close to the eligibility threshold. For instance, this study estimates the effect of universal BCG vaccination for individuals born around April 1st 1975, who are in their mid-forties during the COVID-19 pandemic. Even though some of the BCG protective effects tend to fade rather than increase over time[19], it could still be the case that BCG vaccination at birth has larger effects against COVID-19 on older than middle-aged individuals. When it generalized BCG vaccination in the 1940s, Sweden simultaneously started vaccinating various age groups: newborns in 1943, children entering and leaving school in 1944, and military conscripts in 1945[20]. BCG vaccination rates are not available for cohorts born in the 1930s and 1940s, but it seems likely that the vaccination rate increased gradually, so BCG introduction in Sweden does not lend itself to an RD analysis. There may be other countries where BCG adoption policies in the 1940s and 1950s have created discontinuous changes in the vaccination rate of consecutive cohorts. Studying such policies would be extremely valuable.

Moreover, this study does not measure the COVID-19 immunity benefit conferred by a recent BCG vaccination, as individuals born just before Q2-1975 were vaccinated 45 years ago. The RCTs currently underway will tell if the protective effect of a recent BCG vaccination differs from the effect measured in this study.

Overall, this study shows BCG vaccination at birth does not have a strong protective effect against COVID-19. Thus, it seems that BCG childhood vaccination policies cannot account for the differences in the severity of the pandemic across countries, as had been hypothesized by prior studies[3,5,6]. This advocates for a strict adherence to WHO's recommendation of the vaccine to infants outside of clinical trials[21], and for restraint from starting new clinical trials on this question. The question is of particular importance for a vaccine whose lengthy production process regularly leads to worldwide shortages with dire consequences on children from country with high prevalence of tuberculosis[22].

While RCTs will complement this study by measuring the effect of a recent vaccination, this study comes much before results of the RCTs will be made available, and with a greater precision. Finally, it exemplifies the potential of leveraging past medical policies and statistical techniques designed in the social sciences to answer current medical questions.

**References**

1  Aaby P, Benn CS, Flanagan KL, *et al.* The non-specific and sex-differential effects of vaccines. *Nat Rev Immunol* 2020; published online May 27. DOI:10.1038/s41577-020-0338-x.

2  Moorlag SJCFM, Arts RJW, van Crevel R, Netea MG. Non-specific effects of BCG vaccine on viral infections. *Clin Microbiol Infect* 2019; **25**: 1473–8.

3  Miller A, Reandelar MJ, Fasciglione K, Roumenova V, Li Y, Otazu GH. Correlation between universal BCG vaccination policy and reduced morbidity and mortality for COVID-19: an epidemiological study. *medRxiv* 2020; : 2020.03.24.20042937.

4  Riccò M, Gualerzi G, Ranzieri S, Bragazzi NL. Stop playing with data: there is no sound evidence that Bacille Calmette-Guérin may avoid SARS-CoV-2 infection (for now). *Acta Biomed* 2020; **91**: 207–13.

5   Macedo A, Febra C. Relation between BCG coverage rate and COVID-19 infection worldwide. *Med Hypotheses* 2020; **142**: 109816.

6   Ozdemir C, Kucuksezer UC, Tamay ZU. Is BCG vaccination affecting the spread and severity of COVID-19? *Allergy* 2020; published online April 24. DOI:10.1111/all.14344.

7   Kumar J, Meena J. Demystifying BCG Vaccine and COVID-19 Relationship. *Indian Pediatr* 2020; published online April 30.

8   Lee DS, Lemieux T. Regression Discontinuity Designs in Economics. *Journal of Economic Literature* 2010; **48**: 281–355.

9   COOK TD, WONG VC. Empirical Tests of the Validity of the Regression Discontinuity Design. *Annales d'Économie et de Statistique* 2008; : 127–50.

10  Hyytinen A, Meriläinen J, Saarimaa T, Toivanen O, Tukiainen J. When does regression discontinuity design work? Evidence from random election outcomes. *Quantitative Economics* 2018; **9**: 1019–51.

11  Romanus V, Svensson A, Hallander HO. The impact of changing BCG coverage on tuberculosis incidence in Swedish-born children between 1969 and 1989. *Tuber Lung Dis* 1992; **73**: 150–61.

12  Folkhälsomyndigheten, Veckorapport om covid-19, vecka 20, Tech. rep. Report, May 25th 2020. https://www.folkhalsomyndigheten.se/globalassets/statistik-uppfoljning/smittsamma-sjukdomar/veckorapporter-covid-19/2020/covid-19-veckorapport-vecka-20-final.pdf.

13  Imbens GW, Lemieux T. Regression discontinuity designs: A guide to practice. *Journal of Econometrics* 2008; **142**: 615–35.

14  Calonico S, Cattaneo MD, Titiunik R. Robust Nonparametric Confidence Intervals for Regression-Discontinuity Designs. *Econometrica* 2014; **82**: 2295–326.

15  Calonico S, Cattaneo MD, Titiunik R. Robust Data-Driven Inference in the Regression-Discontinuity Design. *The Stata Journal* 2014; **14**: 909–46.

16  Hollis S, Campbell F. What is meant by intention to treat analysis? Survey of published randomised controlled trials. *BMJ* 1999; **319**: 670–4.

17  Imbens GW, Angrist JD. Identification and Estimation of Local Average Treatment Effects. *Econometrica* 1994; **62**: 467–75.

18  Zwerling A, Behr MA, Verma A, Brewer TF, Menzies D, Pai M. The BCG World Atlas: a database of global BCG vaccination policies and practices. *PLoS Med* 2011; **8**: e1001012.

19  Nguipdop-Djomo P, Heldal E, Rodrigues LC, Abubakar I, Mangtani P. Duration of BCG protection against tuberculosis and change in effectiveness with time since vaccination in Norway: a retrospective population-based cohort study. *Lancet Infect Dis* 2016; **16**: 219–26.

20 Bjartveit K, Waaler H. Some evidence of the efficacy of mass BCG vaccination. *Bull World Health Organ* 1965; **33**: 289–319.

21 Bacille Calmette-Guérin (BCG) vaccination and COVID-19. https://www.who.int/news-room/commentaries/detail/bacille-calmette-guérin-(bcg)-vaccination-and-covid-19 (accessed June 16, 2020).

22 du Preez K, Seddon JA, Schaaf HS, *et al.* Global shortages of BCG vaccine and tuberculous meningitis in children. *Lancet Glob Health* 2019; **7**: e28–9.

**Acknowledgments**

We are grateful to Folkhälsomyndigheten, the Public Health Agency of Sweden, for providing the data used in this study and answering all our questions. We are also grateful to Martin Berlin, Johannes Haushofer, Jerker Jonsson, Ellen Lundqvist, Kyle Meng, Robert Östling, Andrew Oswald, Moa Rehn, and Gonzalo Vazquez-Bare for their help.

**Tables**

Table 1: Effect of BCG-at-birth policy on COVID-19 outcomes, in 1975 Swedish cohort.

| Outcome | Average in 1975 cohort (1) | Effect of BCG vaccination policy (2) | 95% confidence Interval (3) |
|---|---|---|---|
| Cases/1000 inhabitants | 3·230 | -0·002 | [-0·409,0·405] |
| Hospitalizations/1000 inhabitants | 0·758 | 0·089 | [-0·112,0·291] |

*Notes*. This table reports the estimated effect of the BCG-vaccination-at-birth policy on the COVID-19 outcomes of the 1975 cohort in Sweden. The outcomes are the number of cases per 1000 inhabitants and the number of hospitalizations per 1000 inhabitants. For each outcome, its average among the 1975 cohort is shown in Column (1). The estimated effect of the BCG-vaccination-at-birth policy is shown in Column (2), and the 95% confidence interval of this effect is shown in Column (3)

**Figure 1**: Cases per 1000 inhabitants, per quarter of birth

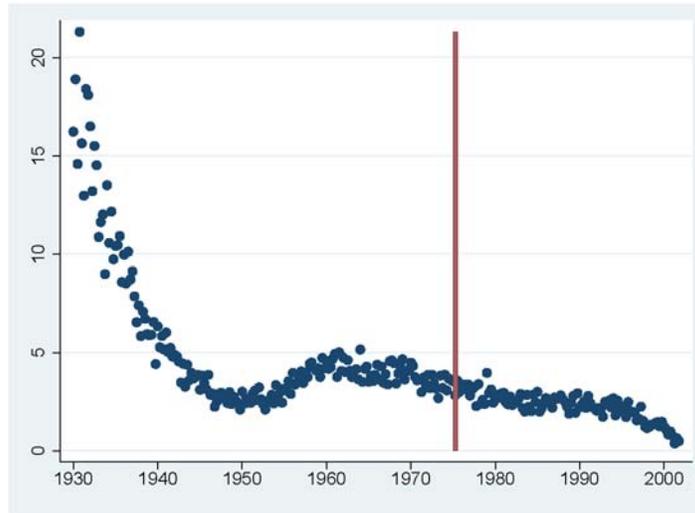

This figure shows the number of COVID-19 cases per 1000 inhabitants per quarter of birth, from Q1-1930 to Q4-2001. Q2-1975, when vaccination at birth was discontinued, is represented by the red vertical line.

**Figure 2**: Hospitalizations per 1000 inhabitants, per quarter of birth

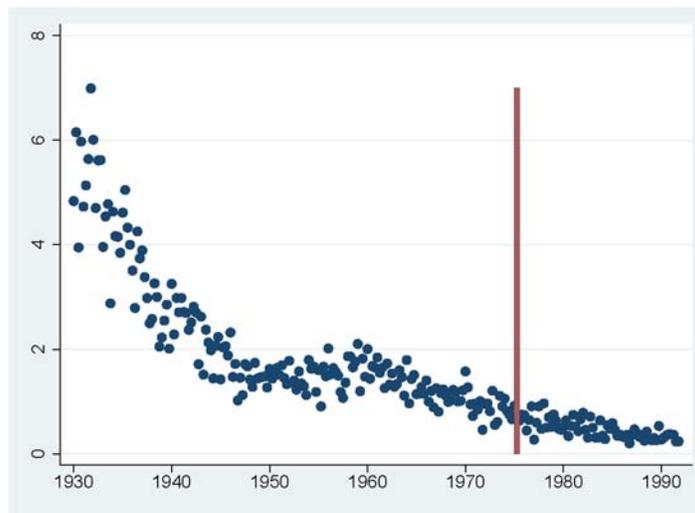

This figure shows the number of COVID-19 hospitalizations per 1000 inhabitants per quarter of birth, from Q1-1930 to Q4-1991. Q2-1975, when vaccination at birth was discontinued, is represented by the red vertical line.

**Figure 3**: Deaths per 1000 inhabitants, per groups of 3 years of birth

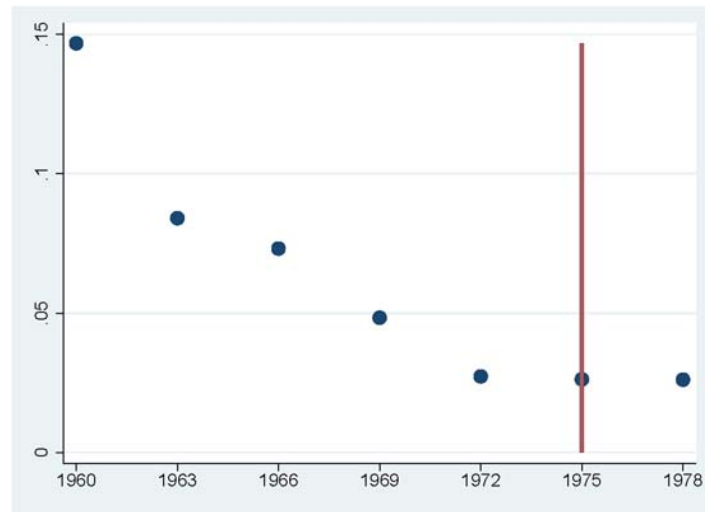

This figure shows the number of COVID-19 deaths per 1000 inhabitants per groups of 3 yearly birth cohorts, from birth cohorts 1960-1961-1962 to birth cohorts 1978-1979-1980. 1975, when vaccination at birth was discontinued, is represented by the red vertical line.

**Supplementary materials**

Table 1

| | | | |
|---|---|---|---|
| Raw variables | 1 | **QBC Cases** | number of reported COVID-19 cases per quarterly birth cohort (QBC), from Q1-1930 to Q4-2001 |
| | 2 | **QBC Hospitalizations** | number of reported COVID-19 cases that led to an hospitalization at diagnosis per QBC, from Q1-1930 to Q4-1991 |
| | 3 | **3-YBC Deaths** | Number of reported COVID-19 deaths, by groups of three yearly birth cohorts (YBC), from YBC 1930-1931-1932 to YBC 1978-1979-1980 |
| | 4 | **YBC Population** | Sweden's population by YBC, from YBC 1930 to YBC 2001, as of December 31st 2019. |
| | 5 | **Percentage of YBC births per QBC** | For each YBC from 1930 to 2001, the percentage of births that took place in Q1, Q2, Q3, and Q4. |
| Constructed variables | 1 | **QBC Population** | for each QBC from Q1-1930 to QBC Q4-2001, the Swedish population in that QBC is defined as YBC Population× the percentage of that year's births that took place in the quarter under consideration (Raw Variable 5) |
| | 2 | **Cases per 1,000 inhabitants** | for each QBC from Q1-1930 to QBC Q4-2001, the number of COVID-19 cases by 1000 inhabitants is defined as QBC Cases/QBC Population×1, 000. |
| | 3 | **Hospitalizations per 1,000 inhabitants:** | for each QBC from Q1-1930 to QBC Q4-1991, the number of COVID-19 hospitalizations at diagnosis by 1000 inhabitants is defined as QBC Hospitalizations/QBC Population×1, 000. |
| | 4 | **Deaths per 1,000 inhabitants** | for each group of three YBCs from YBC 1930-1931-1932 to YBC 1978-1979-1980, the number of COVID-19 deaths by 1000 inhabitants is defined as 3-YBC Deaths/($\sum$ YBC Population), where YBC Population is summed across the three YBCs under consideration. |

Notes: Some QBCs after Q4-1991 had less than 5 hospitalizations, so Folkhälsomyndigheten could not provide their number of hospitalizations due to confidentiality issues.

The Swedish population per QBC is not publicly available, while the population per YBC is. Rather than just dividing the YBC's population by four to recover each QBC's population, we account for the fact that Sweden exhibits a little bit of quarterly birth seasonality. From 1930 to 2001, 25·6%, 27·0%, 24·9%, and 22·5% of births respectively happened during Q1, Q2, Q3, and Q4. To infer a QBC's population, we multiply the population of the corresponding YBC by the proportion this quarter accounts for in the births that took place that year in Sweden. Not all Swedish inhabitants were born in Sweden, so this computation implicitly assumes that quarterly birth seasonality is the same in Sweden as in the countries foreign-born-Sweden-residents immigrated from. Foreign-born individuals only account for 22% of the 1930 to 2001 YBC Swedish population, and quarterly birth seasonality is weak in most countries, so this assumption should not strongly affect the results.